\documentclass[10pt,prl,twocolumn,a4paper,showpacs,superscriptaddress]{revtex4}

\usepackage{graphicx}

\newcommand{\be}{\begin{equation}}
\newcommand{\ee}{\end{equation}}
\newcommand{\bea}{\begin{eqnarray}}
\newcommand{\eea}{\end{eqnarray}}

\newcommand{\f}{\frac}
\newcommand{\vdimer}{{\vrule height0.2cm width0.05cm depth0pt}}
\newcommand{\hdimer}{{\hrule height0.05cm width0.2cm depth0pt}}
\newcommand{\mdimer}{\vbox{\hdimer \vskip 0.0625cm}}
\newcommand{\verdimers}{\hbox{\vdimer \hskip 0.1cm \vdimer}}
\newcommand{\hordimers}{\hbox{\vbox{\hdimer \vskip 0.1cm \hdimer}}}
\newcommand{\dotdimer}{\hbox{$\bullet$ \mdimer}}
\newcommand{\dimerdot}{\hbox{\mdimer \hskip 1.4mm $\bullet$}}

\begin{document}

\title{Continuous-time Diffusion Monte Carlo and the Quantum Dimer Model}

\author{Olav  F. Sylju{\aa}sen}
\affiliation{NORDITA, Blegdamsvej 17, DK-2100 Copenhagen {\O}, Denmark}
\email{sylju@nordita.dk}

%\thanks{Thanks to ...}

\date{\today}

\preprint{NORDITA-2004-72}

\begin{abstract}
A continuous-time formulation of the Diffusion Monte Carlo method for lattice models is presented. In its simplest version, without the explicit use of trial wavefunctions for importance sampling, the method is an excellent tool for investigating quantum lattice models in parameter regions close to generalized Rokhsar-Kivelson points. This is illustrated by showing results for the quantum dimer model on both triangular and square lattices. The potential energy of two test monomers as a function of their separation is computed at zero temperature. The existence of deconfined monomers in the triangular lattice is confirmed.  The method allows also the study of dynamic monomers. A finite fraction of dynamic monomers is found to destroy the confined phase on the square lattice when the hopping parameter increases beyond a finite critical value. The phase boundary between the monomer confined and deconfined phases is obtained.
\end{abstract}

\maketitle
%\section{Introduction}
In the past decade substantial progress have been made in applying Monte Carlo simulations
to study quantum lattice models. Yet there are still quantum models which
are hard to simulate efficiently, even though they do not suffer from the infamous sign problem.
Recently there has been a revived interest in the quantum dimer model (QDM), a quantum model for which it is hard to engineer efficient Monte Carlo updates.

The QDM was first proposed in the context of resonant valence bond (RVB) theories of high-Tc superconductivity\cite{Rokhsar88}. In the RVB theory\cite{Fazekas} pairs of spins form singlets which are approximated by dimers in the QDM.
The RVB scenario is interesting as it implies a gapped spin liquid phase where holes can move freely,
and provides an exotic mechanism for superconductivity\cite{Anderson,Kivelson87}.
Thus a particularly interesting feature of the QDM is the existence of a non-trivial point in parameter space, the Rokhsar-Kivelson (RK) point,
where the model is exactly solvable\cite{Rokhsar88}, and where monomers (holes) are deconfined. However it has been shown by exact studies on small
lattices\cite{Sachdev,Runge}, and repeated here for bigger system sizes, that the RK point
is special and that it is surrounded by crystalline phases such that
monomers are confined for generic values of the
parameters in the QDM on the square lattice. This notion of a deconfined critical point has recently received attention\cite{Moessner01,Deconfined}, and effective field theories describing the nature of small perturbations away from this point has been proposed. The situation on the triangular lattice is different, there it was shown recently\cite{MoessnerSondhi} that the spin liquid behavior exists also away from the RK point.

%For the QDM on the triangular lattice the existence of a spin liquid also
%away from the RK point was shown recently using a finite temperature Monte Carlo method\cite{MoessnerSondhi}. This was based mainly on the absence of structures in the dimer-dimer correlation function at finite temperature. We complement their study by computing the energy of two static monomers as a function of their separation at zero temperature.

\begin{figure}
\includegraphics[clip,width=8cm]{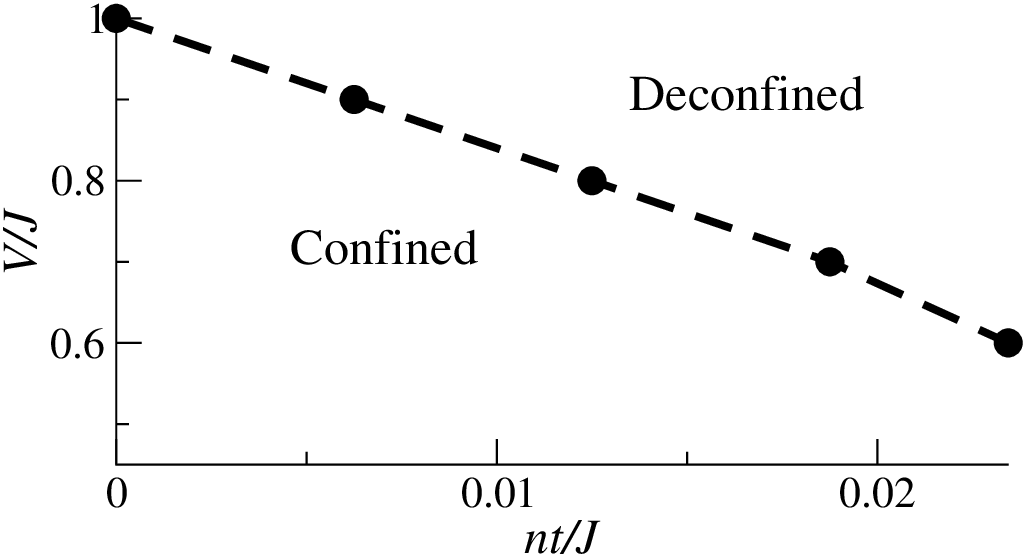}
\caption{The boundary between the monomer confined and deconfined phases in the doped quantum dimer model.\label{phasediagram}}
\end{figure}

While it is interesting
to know how the QDM behaves when the dimers fully cover the lattice, it
is even more important, at least with regards to superconductivity, to know how the model behaves when a finite fraction of monomers is introduced\cite{Moessner01}. The Monte Carlo method introduced here allows
the study of this. We find that the crystalline monomer confining phase is destroyed
when the product of the monomer concentration and hopping increases beyond a finite critical value. Our main result besides the Monte Carlo algorithm itself is the phase boundary between the monomer confined and deconfined phases shown in Fig.~\ref{phasediagram}.

The Hamiltonian of the QDM is
\be \label{Hamiltonian}
	H= -J \sum \left(  \vphantom{\sum} | \verdimers \rangle \langle \hordimers | + \rm{H.c.} \right)
	   +V \sum \left( \vphantom{\sum} | \verdimers \rangle \langle \verdimers | +
                                             | \hordimers \rangle \langle \hordimers |\right)
\ee
where the summations are taken over all elementary plaquettes of the lattice.
The reason why the QDM is a difficult model to simulate stems from
the fact that efficient updates of the classical dimer model are necessarily non-local.
In an imaginary-time formalism a space-time configuration
consists of sheets of dimer configurations
where nearby sheets in the imaginary time direction may differ by the orientation
of two parallel dimers at the same plaquette. While efficient non-local update techniques such
as the directed-loop method\cite{SS} can be utilized for
the classical dimer models\cite{Sandvik03}, just updating the dimers within one
space-time sheet, the QDM seems
a formidable challenge in comparison, as all parallel sheets would need to be updated
simultaneously.
Because of this it is reasonable to expect better performance from Monte Carlo techniques that avoids the constraint imposed by the periodicity in the imaginary-time direction.

Diffusion Monte Carlo (DMC)\cite{Review}
is a genuine zero-temperature technique and contains no references
to the periodicity in the imaginary-time direction.
It is part of a general class of methods known as Projector
Monte Carlo techniques, where a stochastic process is
used to model the result of repeated matrix multiplications.
DMC is the special case where the
iterated matrix is $\exp(-H \Delta \tau)$. Thus it describes, for
infinitesimal $\Delta \tau$, the imaginary-time evolution
of the wavefunction.

As shown by Henley\cite{Henley} the dynamics of the QDM at the RK point can
be obtained using classical continuous-time Monte Carlo.
Thus a continuous-time formulation of the DMC is needed.
It is often stated in the literature that DMC cannot be formulated in continuous (imaginary) time,
and that repeated runs with decreasingly smaller time-intervals must be performed in order to quantify the error induced by a finite time-step.
However, as we will show, a continuous-time formulation of DMC is feasible.
Moreover this method becomes identical to a classical Monte Carlo at the RK point.
The algorithm presented here do not work for spatially continuous systems and one should consult Ref.~\cite{Umrigar} to reduce the time-step errors.

%\section{Algorithm}
The algorithm will be illustrated for a two-state system with Hamiltonian
matrix elements: $H_{ij}= \langle i | H | j \rangle -E_R \delta_{ij}$, where $i,j = \{1,2\}$,
and $E_R$ is a reference energy.
The generalization to larger systems is trivial and is stated in Eq.~(\ref{eq:manystates}).
As in the conventional DMC one introduces $M$ replicas (or walkers) each one representing a basis state of the system. The collection of
replicas represent an instance of the ground state wavefunction
\be
     | \psi \rangle = \left( \begin{array}{c} M_1 \\ M_2 \end{array} \right)
\ee
where $M_j$ is the number of replicas in state $|j\rangle$ ($M_1+M_2 = M$).

We will consider the time evolution operator for an infinitesimal time step $d \tau$. The action of the evolution operator on an instance of the state is
\be \label{eq:evolution}
\left(
\begin{array}{c} M_1^\prime \\ M_2^\prime \end{array} \right)
=
\left(
\begin{array}{cc} 1-H_{11}d \tau & -H_{12} d \tau \\
                   -H_{21}d \tau & 1-H_{22}d \tau
\end{array} \right)
\left(
\begin{array}{c} M_1 \\ M_2 \end{array} \right).
\ee
We will now formulate a stochastic process that on average yields the
evolution equation above:
In the time interval $d \tau$ a replica in
state $|i\rangle$ can undergo one out of four different processes:
(1) with probability $P_{T}(i)$ it can change state to $|j \rangle$, $j \neq i$,
(2) with probability $P_D(i)$ it can ``die'', that is $M_i \to M_i-1$,
(3) with probability $P_R(i)$ it can ``replicate'', that is an extra replica in state $|i \rangle$ is created, and finally
(4) with probability $P_S(i)$ it can stay unchanged in state $|i\rangle$.
As these are {\em all} possibilities,
\be \label{eq:probconservation}
   P_T(i)+P_D(i)+P_R(i)+P_S(i) = 1,
\ee
and must hold for all states $i=1,2$.
Because the off-diagonal matrix element $H_{ji}$ is the only one responsible
for transition between state $i$ and $j$ it is clear that
\be \label{eq:transition}
    P_{T}(i) = -H_{ji} d\tau
\ee
where $j \neq i$. As usual on order to avoid the sign problem off-diagonal
matrix elements are restricted to be negative, which will be assumed in the following.

The increase in number of replicas in state $|i\rangle$ from processes acting on replicas in state $|i\rangle$ is
\be \label{eq:changeindiagreplicas}
  M_i^\prime - M_i = \left[ P_R(i)-P_{T}(i)-P_D(i) \right] M_i.
\ee
This implies when comparing to the diagonal elements of Eq.~(\ref{eq:evolution}) and using Eq.~(\ref{eq:transition})
that
\be \label{eq:replicatingdying}
   P_D(i)-P_R(i) = \left(H_{ii}+H_{ji} \right) d \tau
\ee
where $j \neq i$.
The right hand side of the above takes either a positive or a negative value. We choose $P_R=0$ whenever this value is positive and $P_D=0$ when it is negative. This choice implies that $P_D$ and $P_R$ are of the order $d \tau$ as also holds for $P_T$. The probability conservation equation Eq.~(\ref{eq:probconservation}) then implies
\be \label{eq:staying}
  P_S(i) = 1 -  \left( |  H_{ii}+H_{ji} | - H_{ji} \right)  d\tau
\ee
Thus for most time intervals nothing happens to each replica.
This is analogous to stochastic modeling of the radioactive decay problem,
although here with three different decay channels\cite{BeardWiese}. Thus we can simulate
the imaginary-time evolution of one replica by generating exponentially
distributed decay times with decay constant $|H_{ii}+H_{ji}|-H_{ji}$.
Having obtained the decay time, the type of decay is determined
stochastically proportionally to the respective probabilities $P_T, P_D$ and $P_R$.

The generalization to larger systems is immediate. In Eq.~(\ref{eq:changeindiagreplicas}) $P_T(i)$ should be changed to account for transitions to all possible states different from $|i\rangle$.
It follows that the only change to Eqs.~(\ref{eq:replicatingdying}) and (\ref{eq:staying}) is to set
\be \label{eq:manystates}
H_{ji}   \to \sum_{j \neq i} H_{ji}.
\ee

In an actual simulation each replica contains information about the state of the system as well as a ``clock''
indicating the starting time for the next evolution. The replicas are ordered in a list.
They all start in the same state and with their clock set to 0.
Each replica in the list is subsequently evolved up to a control time $\tau_c$,
or until the replica dies in which case it is removed from the list. An evolution of a replica
begins by generating a decay-time $\tau_d$ according to the exponential distribution. If
$\tau_d > \tau_c$ the clock is set to $\tau_c$ and evolution of the next replica starts.
If however $\tau_d < \tau_c$, the clock is set to $\tau_d$ and a random number is drawn to
select the decay type. If the decay type is (1) the state of the replica changes, if
it is (2) the replica is removed from the list and
if it is (3) a copy of the replica with clock set to $\tau_d$ is inserted at the end of the list.
As long as the replica is not dead the evolution continues by picking a new decay time until $\tau_c$ is reached.
When the last replica in the list has evolved up to $\tau_c$, all replicas have the same clock-time, and
measurements can be performed. The process is repeated by increasing $\tau_c$ and starting over from the beginning of the replica list.

The control times are included in order to perform population control to avoid an explosion/implosion in the number of replicas. Population control is achieved by changing the value of the reference energy $E_R$ so as to maintain a roughly constant number of replicas.
While a constant $E_R$ is innocuous, a time-varying one is not. Thus the simulation with population control is not simulating exactly the time evolution operator. A way to correct this is to reweight the simulations by
a factor that corrects for the time-varying reference energy\cite{Umrigar}.

It is known that importance sampling reduces statistical errors in DMC. Importance
sampling is achieved by sampling the product of the wavefunction times a trial wavefunction
instead of the wavefunction alone. The trial wavefunction is chosen such that branching due to the replicating and death processes is minimized. We note that according to Eq.~(\ref{eq:replicatingdying})
branching is {\em absent} when the potential energy of a state equals the kinetic energy associated with motion away from the same state. This is exactly the condition defining generalized RK points\cite{Henley}.
Thus no explicit trial wavefunction is needed at RK points. Formally the (implicit) trial wavefunction
is simply 1, the equal superposition of all basis states, which in fact is the exact ground-state wavefunction. Thus we expect small statistical errors in the vicinity of RK points even without introducing
explicit trial wavefunctions.

\begin{figure}
\includegraphics[clip,width=8cm]{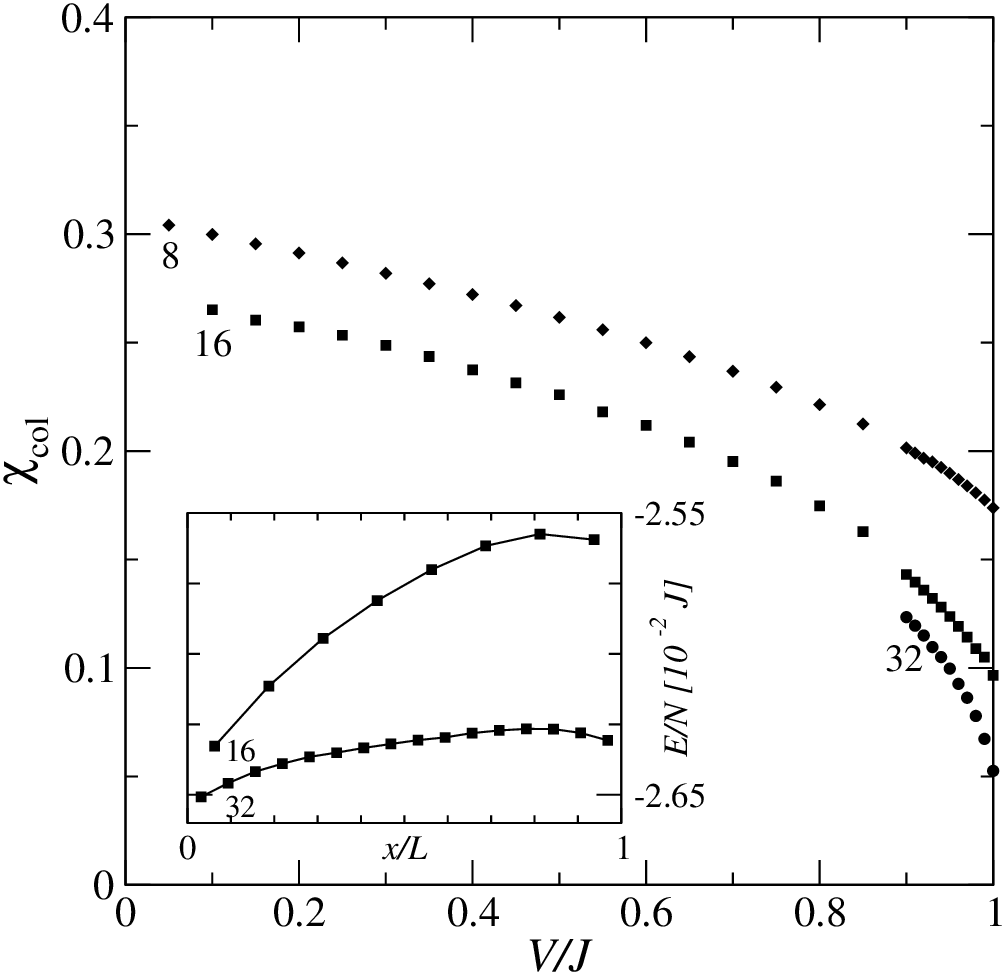}
\caption{Columnar order parameter as a function of $V/J$ on square lattices for different linear lattice sizes $L$(labels on each curve). The inset shows the energy per site as a function of monomer separation for $V/J=0.9$. Error bars are smaller than the symbol sizes.\label{columnar}}
\end{figure}

%\section{Quantum Dimer model}
We now turn to simulations of the QDM.
First we calculate the columnar order parameter, $\chi_{\rm col}$, defined by
\be
\chi_{\rm col}^2 = \f{1}{4N^2} \langle 
\left( \sum n_{\rm H}(\vec{r}) (-1)^{r_x} \right)^2
+
\left( \sum n_{\rm V}(\vec{r}) (-1)^{r_y} \right)^2
\rangle
\ee
where $n_{\rm H}$ ($n_{\rm V}$) is the number of horizontal (vertical) dimers belonging to the plaquette at $\vec{r}$. $\vec{r}$ is an integer-valued coordinate labelling the centre of each plaquette on a lattice of size $N=L \times L$. The sums are to be taken over all plaquettes.
%as defined in Eq.~(4)  of Ref.\cite{Runge} on a square
%lattice of size $N= L \times L$. 
The results are shown in
Fig.~\ref{columnar} and are carried out for significantly larger lattice sizes
than in the exact studies Refs.~\cite{Sachdev,Runge}.
In all simulations presented here $M=1000$ replicas were used. Diagonal observables were measured using the forward-walking technique\cite{forward} keeping histories for times typically of the order 100/J. Energies
were measured using the standard growth estimator\cite{Umrigar}. To correct for the time-varying $E_R$ we kept track
of $E_R$ for typical times $200/J$ and used the approach described in Ref.~\cite{Umrigar} to reweight
the simulation.

Extrapolating the columnar order parameter to infinite lattice size we find a linear behavior
in $1/L$ at the RK point extrapolating to zero within error bars, and $1/L^2$ corrections for $V<J$ extrapolating to finite values. Our error bars and system sizes investigated strictly allows the conclusion that the order parameter is finite for $V<0.99J$, although we believe it is finite for all $V<J$.

The existence of a finite columnar order parameter implies that monomers
are confined. In order to show this explicitly we also compute the energy
of two static monomers at different separations.
Two monomers with a horizontal separation $x$ is inserted in a configuration with horizontal columnar order.
The insertion causes the dimers directly between the monomers to be displaced one lattice spacing.
At $V=J$ the energy is independent of separation, but for $V<J$ the energy increases as the separation between monomers is increased. This is shown in the inset of Fig.~\ref{columnar} for $V/J=0.9$ and implies the confinement of monomers. Naively one would expect the graphs to be symmetric around $L/2$ because of the periodic boundary condition. However for small square lattices with $L$ even the conservation of winding numbers prohibits this symmetry.
\begin{figure}
\includegraphics[clip,width=8cm]{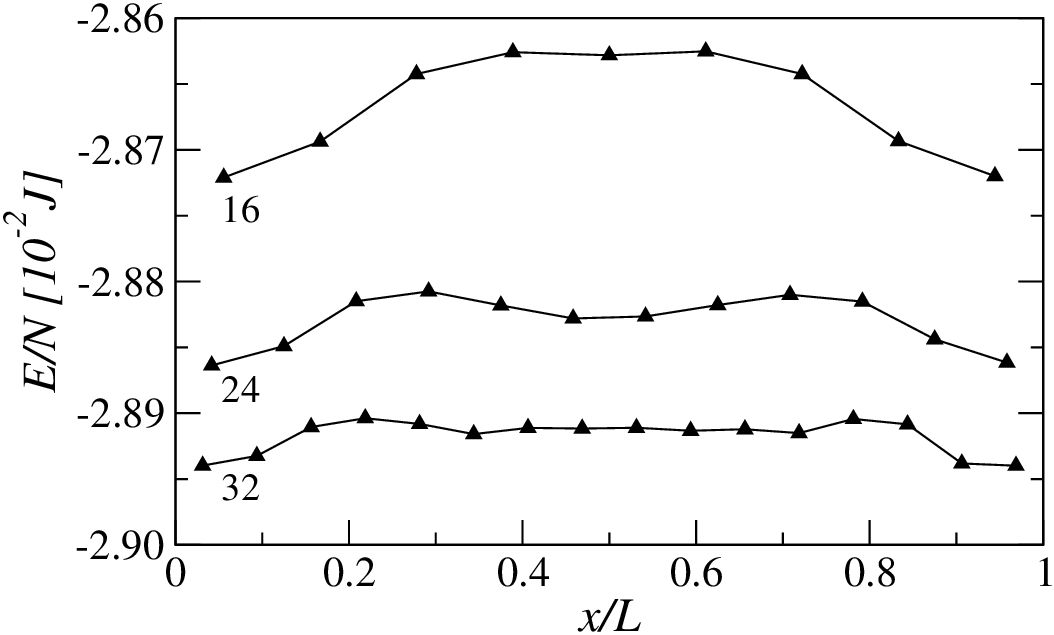}
\caption{Energy per site as functions of monomer separation for different system sizes $L$(labels on each curve) on the triangular lattice.\label{confinement}}
\end{figure}

Fig.~\ref{confinement} shows the energy of two static monomers on the {\em triangular} lattice as a function of their separation for $V/J=0.9$. The
contrast with the square lattice case is evident. On the triangular lattice the energy increases for a few lattice spacings and then drops slightly reaching a plateau. Thus there is no confining potential. This agrees with the conclusions reached in Ref.\cite{MoessnerSondhi}.

We now turn to the study of monomers with their own dynamics. Specifically the term\cite{Rokhsar88}
\be
  H_{\rm dyn} = -t \sum_{\langle ij \rangle, \langle jk \rangle}
  \left( \vphantom{\sum} | \dotdimer \rangle \langle \dimerdot | + \rm{H.c.}  \right)
\ee
is added to the Hamiltonian Eq.~(\ref{Hamiltonian}). The sum is taken over triplet of sites $i,j$ and $k$ such that both $i$ and $j$, and $j$ and $k$ are nearest neighbors.

\begin{figure}
\includegraphics[clip,width=8cm]{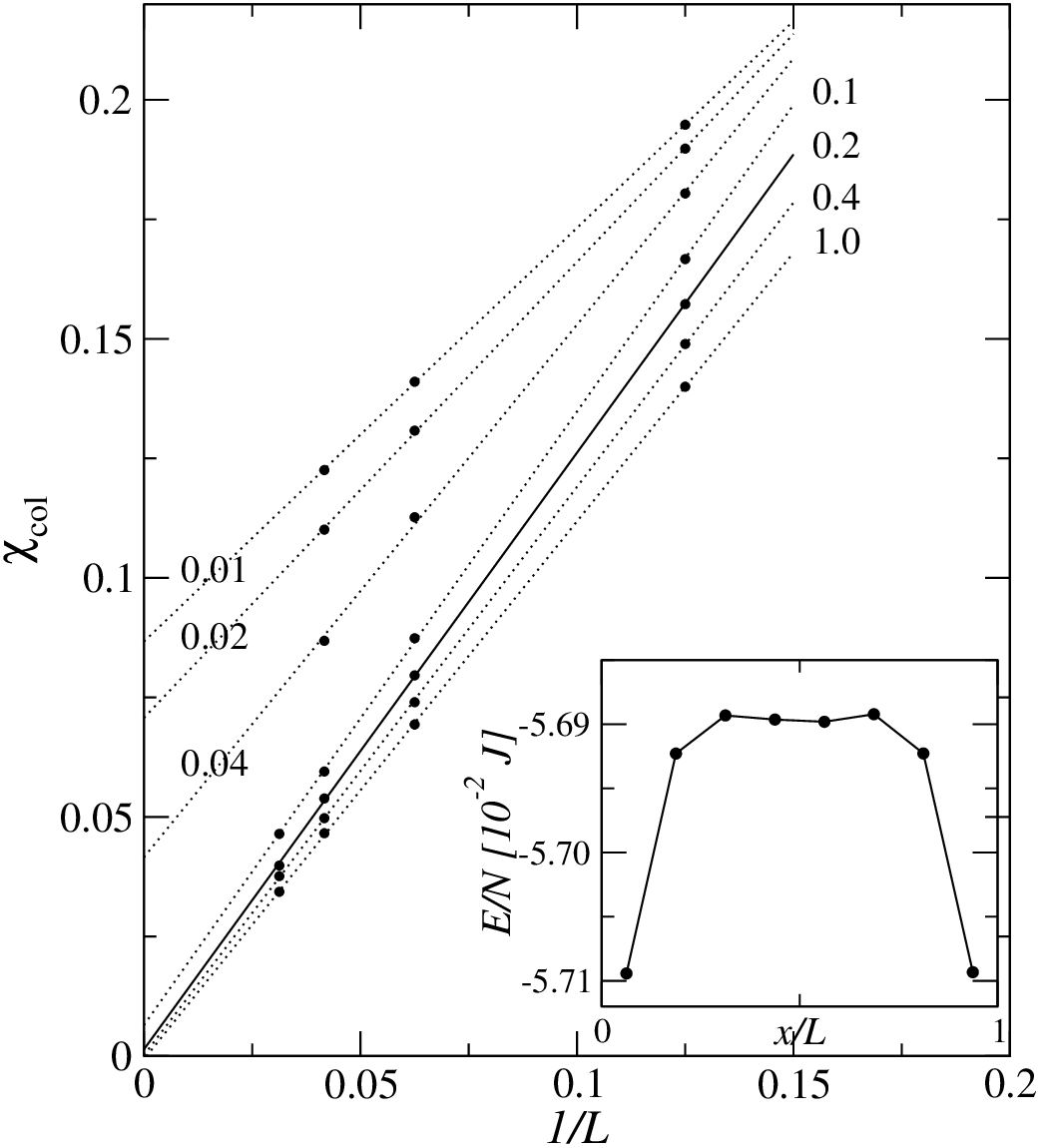}
\caption{Columnar order parameter on square lattices with dynamic monomer concentration $n=1/32$ and $V/J=0.9$ as a function of $1/L$ for different values of $t/J$(labels on each line). The lines are least square linear fits to the data points.
The inset on the lower right shows energy versus separation of two static monomers in a background of dynamic monomers and dimers for $n=1/8$, $L=16$, $V/J=0.9$  and $t/J=0.1$.  \label{dynamic}}
\end{figure}

%Consider first two dynamic holes for different lattice sizes. For definiteness we hereafter fix $V/J=0.9$.
%We find that the columnar order parameter drops rapidly for small $t < 0.1J$, but decreases only slowly for larger $t$. Fixing $t$ and performing a finite size scaling we find that the small $t$ points scales as $1/L$ towards a value $0.1$ while the higher $t$ points scales as $1/L^2$ towards roughly the same value. Thus two dynamic holes do not destroy the columnar order in the infinite system.
%For a finite fraction $n$ of dynamic holes the situation is different. 
In Fig.~\ref{dynamic} we plot the columnar order parameter as a function of $1/L$ for different values of $t$ for a small monomer fraction away from the RK-point.
For $t>t_c \approx 0.2J$ the columnar order extrapolates linearly to zero as $L \to \infty$. For
smaller values of $t$ this order parameter is finite. Thus there exist a finite critical $t_c$ where
the columnar order is destroyed. Increasing $n$ to $1/16$, and $1/8$ we find
that $t_c$ drops inversely proportional to $n$. 
A computation of the energy of two static monomers in a background of dimers and dynamic monomers for $t>t_c$ is shown in the lower right inset of Fig.~\ref{dynamic}. This indicates that the dynamic monomers screen the interaction between monomers, rendering the potential deconfining.
Calculating the values of $t_c$ for other values of $V/J$ we obtain the phase boundary between the confined and deconfined phases shown in Fig.~\ref{phasediagram}

The method presented here is applicable to any quantum lattice models that do not suffer from the sign problem. This includes
quantum vertex models\cite{Chakravarty} and bosonic models with internal degrees of freedom.
\begin{acknowledgments}
Monte Carlo calculations were in part carried out using NorduGrid,
a Nordic facility for Wide Area Computing and Data Handling.
\end{acknowledgments}


\begin{thebibliography}{00}
\bibitem{Rokhsar88} D. S. Rokhsar and S. A. Kivelson, Phys. Rev. Lett. {\bf 61}, 2376 (1988).
\bibitem{Fazekas}P. W. Anderson, Mater. Res. Bull. {\bf 8}, 153 (1973); P. Fazekas and P. W. Anderson, Philos. Mag. {\bf 30}, 23 (1974).
\bibitem{Anderson} P. W. Anderson, Science, {\bf 235}, 1196 (1987).
\bibitem{Kivelson87} S. A. Kivelson, D. S. Rokhsar and J. P. Sethna, Phys. Rev. B {\bf 35}, R8865 (1987).
\bibitem{Sachdev} S. Sachdev, Phys. Rev. B {\bf 40}, 5204 (1989).
\bibitem{Runge} P. W. Leung, K. C. Chiu and K. J. Runge, Phys. Rev. B {\bf 54}, 12938 (1996).
\bibitem{Moessner01} R. Moessner, S. L. Sondhi and E. Fradkin, Phys. Rev. B, {\bf 65}, 024504 (2001).
\bibitem{Deconfined} E. Fradkin, D. A. Huse, R. Moessner, V. Oganesyan and S. L. Sondhi, Phys. Rev. B {\bf 69}, 224415 (2004);
A. Vishwanath, L. Balents, and T. Senthil, Phys. Rev. B {\bf 69}, 224416 (2004);
E. Ardonne, P. Fendley and E. Fradkin, Annals Phys. {\bf 310}, 493 (2004).
\bibitem{MoessnerSondhi} R. Moessner and S. L. Sondhi, Phys. Rev. Lett. {\bf 86}, 1881 (2001).
\bibitem{SS} O. F. Sylju{\aa}sen and A. W. Sandvik, Phys. Rev. E, {\bf 66}, 046701, (2002).
\bibitem{Sandvik03} A. W. Sandvik, [arXiv: cond-mat/0312097].
\bibitem{Review} for a review and further refs. see D. Ceperley, B. Alder, Science, {\bf 231}, 555 (1986).
\bibitem{Henley} C. L. Henley, J. Phys.: Condens. Matt. {\bf 16}, S891 (2004)
\bibitem{Umrigar} C. J. Umrigar, M. P. Nightingale, and K. J. Runge, J. Chem. Phys. {\bf 99}, 2865 (1993).
\bibitem{BeardWiese} This observation was also employed to formulate world-line algorithms in continuous time, B. B. Beard and U.-J. Wiese, Phys. Rev. Lett. {\bf 77}, 5130 (1996).
\bibitem{forward} K.S. Liu, M.H. Kalos, and G.V. Chester, Phys. Rev. A {\bf 10}, 303 (1974).
\bibitem{Chakravarty} S. Chakravarty, Phys. Rev. B {\bf 66}, 224505 (2002).











\end{thebibliography}
\end{document}